\documentclass{ws-procs9x6-cpt16}

\begin{document}

\newcommand{\refeq}[1]{(\ref{#1})}
\def\etal {{\it et al.}}

\def\mbf#1{\mbox{\boldmath$#1$}}
\def\cb{\overline{c}}
\def\cbw{\cb^w}
\def\mn{{\mu\nu}}
\def\ab{\overline{a}}
\def\abw{\ab^w}
\def\sb{\overline{s}}

\title{Prospects for SME Tests with Experiments at SYRTE and LKB}

\author{C.\ Guerlin,$^{1,2}$ H.\ Pihan-Le Bars,$^{2}$ 
Q.G.\ Bailey,$^{3}$ and P.\ Wolf$^2$}

\address{$^1$Laboratoire Kastler Brossel, ENS-PSL Research University, CNRS, \\ 
UPMC-Sorbonne Universit\'es, Coll\`ege de France,
75005 Paris, France}

\address{$^2$SYRTE, Observatoire de Paris, PSL Research University, CNRS, \\ 
Sorbonne Universit\'es, UPMC, 75014 Paris, France}

\address{$^3$Embry-Riddle Aeronautical University,
Prescott, Arizona 86301, USA}

\begin{abstract}
Preliminary work has been done 
in order to assess the perspectives of metrology and fundamental physics
atomic experiments at SYRTE and LKB in the search for physics beyond
the Standard Model and General Relativity. The first
studies we identified are currently ongoing with the
Microscope mission and with a Cs
fountain clock. The latter brings significant improvement on
the proton-sector coefficient $\cb_{TT}$ down to the $10^{-17}$ GeV level.
\end{abstract}

\bodymatter

\section{Experiments and SME sectors of interest}\label{sec1}
We investigate the use of atomic experiments and Earth orbit space
missions at SYRTE and LKB in the search for Lorentz-invariance
violations (LIV) in the framework of the Standard-Model Extension (SME). 
We worked, mainly in the minimal SME, on
gathering from the literature LIV effects, models, and perspectives for
the experiments considered (Secs.\ \ref{sec2} and \ref{sec3}). We identified the most promising and feasible
experiments and data analysis, which is currently work in progress
(Sec.\ \ref{sec4}). The
experiments considered include: the double-species Cs fountain clock (FO2),
the cold atom gravimeter (CAG), the superconducting gravimeter, the GBAR
experiment,\cite{Perez2012} the ACES mission,\cite{Laurent2015}
and the Microscope Weak Equivalence Principle (WEP) test mission.\cite{Touboul2002}
The SME models for these
experiments involve the fermion, gravity, and photon
sectors. We focused on the matter CPT-even $\cbw_\mn$ tensor
and CPT-odd $\abw_\mu$ vector
($w=e,p,n$), on disentangling them, and on pushing constraints towards or beyond their expected Planck scale
suppression, i.e., $10^{-19}$ GeV.

\section{SME LIV in center of mass motion}\label{sec2}

Including gravitational-sector LIV with matter-gravity couplings
in a classical treatment modifies the classical dynamics of massive bodies. As shown in
Eq.\ (132)-(134) in Ref. \refcite{Kostelecky2011},
the effective inertial mass of test bodies gets tensor components due to $\cbw_\mn$, while
$\cbw_{\mathrm{00}}$ and $\abw_{\mathrm{0}}$ modify the gravitational
force. It results in isotropic and frame-dependent LIV modifications
to the effective acceleration. The isotropic part is $g^T=g(1+\beta^T)$ with
$\beta^T=\tfrac{2\alpha}{m^T}(\bar a^T)_0-\tfrac{2}{3}(\bar c^T)_{00}$ for
a test mass $m^T$. Due to a change of the $\abw_0$ sign for antimatter, 
free-fall LIV could still be large for antimatter while strongly
constrained for normal matter,\cite{Tasson2014} as can be measured,
e.g., in GBAR in the future (free fall of antihydrogen).\cite{Perez2012} Deriving the equations of motion from the lagrangian by the Euler-Lagrange
equations and expressing frame dependence with respect to a common
inertial frame (the Sun-centered celestial equatorial frame) leads to the full LIV time variation model, which we can 
use to model the observables of gravimeters and WEP 
tests.\cite{Kostelecky2011} Note that for any neutral test body,
the proton to electron SME coefficient ratio is fixed in this type of test.

\section{SME LIV in atomic internal energy}\label{sec3}

Atoms are composite systems in quantized
bound states. SME LIV energy shifts can be calculated perturbatively
from the nonrelativistic free-fermion hamiltonian. Spin-independent
contributions can be classified in LIV and metric
fluctuation order as\cite{Kostelecky2011} $ H=H^{(0,1)}+H^{(1,1)}_c$.

The first term describes LIV shifts for clocks in flat
spacetime,\cite{BluhmLane} arising
as an anisotropy of the fermionic dispersion relation. Indeed nonzero
LIV expectation
values arise for each particle from the operator $\delta H=-{\cal C}^{(2)}_0 {\cal
  P}^{(2)}_0/6m$ (in spherical tensor notation) which involves
the quadrupole moment of $\cbw_\mn$ and $p_ip_j$ tensors: ${\cal
  C}^{(2)}_0=\cb_{jj}-3\cb_{33}=\cb_Q/m$ and ${\cal P}^{(2)}_0={\mbf p}^2-3p_3^2$
in cartesian coordinates with the laboratory frame axis $3$ along the
quantization axis. 
Analysis of sidereal variations in magnetized Zeeman substates,
either by measuring hyperfine
frequency in a Cs fountain clock\cite{Wolf2006} or Larmor
precession in co-magnetometers,\cite{Smiciklas2011} has allowed constraining proton
and neutron anisotropic $\cbw_\mn$ coefficients beyond the Planck scale suppression. The analysis is in fact nucleus model dependent, which is
currently being investigated.\cite{BrownFlambaum} For electrons, the best sensitivity
has been reached with dysprosium\cite{Hohensee2013} and
Ca$^+$ ions\cite{Pruttivarasin2015} spectroscopy.

The second term describes LIV modification in curved spacetime.  It is
smaller since of overall second order but
relevant for constraining the isotropic $\cbw_{TT}$
coefficients, otherwise suppressed by boosts and often dismissed in
analyses. This term is $\delta H=(2U/3c^2)\cbw_{00}\mbf {p}^2/2m$
with $U$ the newtonian gravitational potential.
It can be seen as a potential-dependent rescaling of electron
and proton inertial masses, leading to modified binding energies and thus to a modified
clock gravitational redshift. Redshift measurement on
clocks as well as
null redshift tests, in the gravitational field of Earth
and Sun, thus
constrain proton and electron $\cbw_{TT}$
coefficients.\cite{Kostelecky2011,Hohensee2011} This brings a competitive limit for the latter one at $10^{-8}$ GeV.\cite{Hohensee2013}

Note that within the SME, WEP
violations and modified reshift come together,
both due to species dependent LIV. In addition, WEP violation is also
slightly modified by internal energy LIV. Indeed nuclear binding energy
contributes to up to 1$\%$
of the atom's rest mass, so LIV modification to nuclear binding energy
also modifies the
free fall of the atom.\cite{Hohensee2011,Hohensee2013a} 
This helps disentangling electron from
proton contributions.

\section{Progress and perspectives}\label{sec4}
For gravimeter experiments, 
our progress so far concerns the model and
systematics evaluation. 
Combining Refs.\ \refcite{Kostelecky2011} and \refcite{Bailey2006}, 
we derived the
gravimeter model including the Earth's spherical inertia $i_E$,
$\cbw_{\mn}$, $\abw_{\mu}$ and $\sb_{\mn}$, to order $V_L,V_E$ (laboratory
and Earth boost). This improved form has been used
in Ref.\ \refcite{Hees2015} in Eq. (21). 
We then questioned whether tidal models used in
gravimeters subtract SME
signatures. Indeed tidal models are usually fitted to data related to
free fall, while all SME frequencies overlap with tidal frequencies.\cite{Tamura1987}  To our knowledge this issue has not 
been addressed yet in detail in the literature for gravimeter tests. We are currently working along 
this direction.
Progress of the performances of atom gravimeters since the analysis in Ref.~\refcite{Chung2009} could give interesting improved constraints, as well as analyzing the longer time
series that are available from geophysics observations (this approach was
independently mentioned and a preliminary discussion
given by J.D.\ Tasson.\cite{Tasson2016})

For the spin-polarized Cs fountain clock,
keeping the second order boost suppressed terms for all $\cb_\mn$ coefficients, we
reanalyzed the data presented in Ref.\ \refcite{Wolf2006}. We made only few
assumptions in our model and treatment, keeping track of all
correlations present in our data, and calculated confidence intervals.
This allowed us to constrain the  proton $\cb_{TT}$ coefficient, in the
Schmidt model, down to the $10^{-17}$ GeV scale, improving present limits by
six orders of magnitude compared to WEP tests,\cite{Kostelecky2011,datatables} bringing it close to the Planck scale
suppression. More details will be given in a dedicated article.\cite{HPB}

The Microscope mission was launched on April 25, 2016,
aiming for improving the WEP test\cite{Touboul2002} down to $10^{-15}$. Based on
Ref.\ \refcite{Kostelecky2011}, a proposal for SME
analysis 
has been
accepted by CNES and the Science Working Group. We are currently
investigating the calibration procedure of systematic effects in order to
identify at which data treatment level we should proceed with an SME
analysis without losing possible LIV signals.  We expect this mission
to set the best limits on all $\abw_\mu$ coefficients down to
$10^{-13}$ GeV (isotropic coefficient), improving
present constraints by three to six orders of magnitude.\cite{Hees2015,datatables}

Beyond these tests, we are performing simulations of spin polarized states LIV on ACES.\cite{Laurent2015}
Work on nuclear models is also being pursued. 

\section*{Acknowledgments}
C.G., Q.G.B., and P.W.\ acknowledge support from the Sorbonne Universit\'es
grant Emergence for the CABESTAN collaboration. Q.G.B.\ acknowledges
the National Science Foundation under grant number PHY-1402890.


\begin{thebibliography}{xx}

\bibitem{Perez2012}
P.\ Perez and Y.\ Sacquin, 
Class.\ Quantum Grav.\ {\bf 29}, 184008 (2012).

\bibitem{Laurent2015}
P.\ Laurent \etal, 
C.R.\ Physique {\bf 16}, 540 (2015).

\bibitem{Touboul2002}
P.\ Touboul \etal, 
Acta Astronautica {\bf 50}, 433 (2002).

\bibitem{Kostelecky2011}
V.A.\ Kosteleck\'y and J.D.\ Tasson, 
Phys.\ Rev.\ D {\bf 83}, 016013 (2011).

\bibitem{Tasson2014}
J.D.\ Tasson, 
Hyperfine Int.\ {\bf 228}, 111 (2014).

\bibitem{BluhmLane}
V.A.\ Kosteleck\'y and C.D.\ Lane,
Phys.\ Rev.\ D {\bf 60}, 116010 (1999); 
R. Bluhm \etal,
Phys.\ Rev.\ D {\bf 68}, 125008 (2003).

\bibitem{Wolf2006}
P.\ Wolf \etal,
Phys.\ Rev.\ Lett.\ {\bf 96}, 060801 (2006).

\bibitem{Smiciklas2011}
M.\ Smiciklas \etal, 
Phys.\ Rev.\ Lett.\ {\bf 107}, 171604 (2011).

\bibitem{BrownFlambaum}
B.A.\ Brown \etal,
arXiv:1604.08187;
V.V.\ Flambaum, 
arXiv:1603.05753.

\bibitem{Hohensee2013}
M.A.\ Hohensee \etal,
Phys.\ Rev.\ Lett.\ {\bf 111}, 050401 (2013).

\bibitem{Pruttivarasin2015}
T.\ Pruttivarasin \etal,
Nature {\bf 517}, 592 (2015).

\bibitem{Hohensee2011}
M.A.\ Hohensee \etal,
Phys.\ Rev.\ Lett.\ {\bf 106}, 151102 (2011).

\bibitem{Hohensee2013a}
M.A.\ Hohensee \etal,
Phys.\ Rev.\ Lett.\ {\bf 111}, 151102 (2013).

\bibitem{Bailey2006}
Q.G.\ Bailey and V.A.\ Kosteleck\'y, 
Phys.\ Rev.\ D {\bf 74}, 045001 (2006).

\bibitem{Hees2015}
A.\ Hees \etal,
Phys.\ Rev.\ D {\bf 92}, 064049 (2015).

\bibitem{Tamura1987}
Y.\ Tamura, 
Bull.\ Inf.\ Mar{\'e}es Terrestres {\bf 99}, 6813 (1987).

\bibitem{Chung2009}
K.Y.\ Chung \etal, 
Phys.\ Rev.\ D {\bf 80}, 016002 (2009).

\bibitem{Tasson2016}
J.D.\ Tasson, 
these proceedings.

\bibitem{datatables}
{\it Data Tables for Lorentz and CPT Violation,}
V.A.\ Kosteleck\'y and N.\ Russell,
2016 edition,
arXiv:0801.0287v9.

\bibitem{HPB}
H.\ Pihan-Le Bars \etal, 
in preparation.

\end{thebibliography}
\end{document}